\documentclass[10pt,aps,prl,twocolumn,superscriptaddress]{revtex4}

\usepackage[utf8]{inputenc}
\usepackage[linktocpage=true, colorlinks=true, urlcolor=blue, linkcolor=blue, citecolor=blue]{hyperref}
\usepackage{graphicx}
\usepackage{amsmath,amssymb}
\usepackage{color}
\usepackage{subfigure}
\usepackage{url}

\graphicspath{{Images/}}
\begin{document}

\title{Holographic imaging of surface acoustic waves}

\author{Fran\c{}cois Bruno}
\author{J\'e{}r\^o{}me Laurent}
\author{Daniel Royer}
\author{Michael Atlan}

\affiliation{
Institut Langevin. Fondation Pierre-Gilles de Gennes. Centre National de la Recherche Scientifique (CNRS) UMR 7587, Institut National de la Sant\'e et de la Recherche M\'edicale (INSERM) U 979, Universit\'e Pierre et Marie Curie (UPMC), Universit\'e Paris 7. \'Ecole Sup\'erieure de Physique et de Chimie Industrielles ESPCI ParisTech - 1 rue Jussieu. 75005 Paris. France\\
}

\date{\today}

\begin{abstract}
We report on an experimental demonstration of surface acoustic waves monitoring on a thin metal plate with heterodyne optical holography. Narrowband imaging of local optical pathlength modulation is achieved with a frequency-tunable time-averaged laser Doppler holographic imaging scheme on a sensor array, at video-rate. This method enables robust and quantitative mapping of out-of-plane vibrations of nanometric amplitudes at radiofrequencies.
\end{abstract}

\maketitle

Imaging acoustic fields at the surface of solids is of primary importance in non destructive testing applications. Laser Doppler interferometric methods are commonly used for non-contact measurements of surface acoustic waves in particular \cite{WhitmanKorpel1969, DeLaRue1972, Stegeman1976}. These methods exhibit high reliability either in wideband \cite{RoyerDieulesaint1986, RoyerDieulesaint1989, JiaBoumiz1993} or narrowband \cite{DeLaRue1972, Monchalin1985, BramhavarPouet2009} single point vibration measurements. Wideband methods allow for transient vibration sensing \cite{BarriereRoyer2001} with a temporal resolution given by the reciprocal of sensor bandwidth, while narrowband schemes permit high frequency resolution and better noise-limited sensitivity with respect to wideband approaches. Absolute measurement of the optical pathlength modulation depth (and hence the out-of-plane vibration amplitude) can be readily derived from the ratio of the first optical sidebands' magnitude to the non-shifted optical carrier magnitude for narrowband measurements \cite{WhitmanKorpel1969}. These sidebands appear with phase modulation of optical waves, as a result of bouncing onto a surface in sinusoidal motion. Laser phase modulation has been extensively used to measure vibrations in nondestructive testing applications. However, imaging requires time-consuming scanning of the tested sample \cite{Rembe2006}. Wide-field imaging of surface acoustic waves of nanometer amplitude can be achieved by optical holographic  arrangements, either in homodyne \cite{Powell1965, PicartLeval2003, Pedrini2006} or heterodyne \cite{Aleksoff1971, UedaMiida1976, JoudLaloe2009} recording conditions. In particular, the off-axis configuration enables reliable imaging of out-of-plane mechanical vibrations because unwanted interferometric contributions can be filtered-off, and high sensitivity ensured. Nevertheless, quantitative measurements of vibration amplitudes much smaller than the optical wavelength with an array detector remain difficult to achieve. The measurement of vibration amplitudes down to a few Angstroms was linked to the signal-to-noise ratio of the detection, for single sideband modulation of the optical local oscillator (LO) \cite{CumminsKnable1963}, in the absence of spurious effects \cite{Aleksoff1971, UedaMiida1976, Togami1978}. Later on, nanometric vibration amplitude measurements were achieved with digital holography, by sequential measurements of the first optical sideband and the non-shifted light component \cite{PsotaLedl2012} and by simultaneous measurements in frequency-division multiplexing regime \cite{VerrierAtlan2013}.

In this letter, we report on the use of holographic detection as a multimode coherent optical detection for steady-state surface acoustic waves measurements. This method benefits from shot-noise limited sensitivity \cite{GrossAtlan2007} as  single mode schemes \cite{WhitmanKorpel1969}. It also benefits from a large optical \'e{}tendue \cite{GrossDunn2005}, which makes it suitable for wide-field imaging of out-of-plane vibrations with noise-limited amplitude resolution. In the presented method, scanning of the probe laser beam is unnecessary: the sample is illuminated in wide field. Steady-state surface acoustic waves monitoring in a thin metal plate and calibration of the holographic probe against a single mode heterodyne laser probe are presented.

\begin{figure}[]
\centering
\includegraphics[width = 8.0 cm]{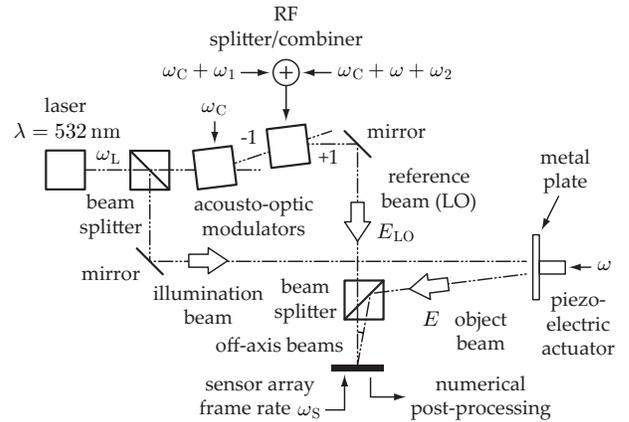}
\caption{Sketch of the Mach-Zehnder holographic interferometer designed for surface acoustic wave imaging.}
\label{fig_SetupSurfaceAcousticWaves}
\end{figure} 

\begin{figure}[]
\centering
\includegraphics[width = 7.0 cm]{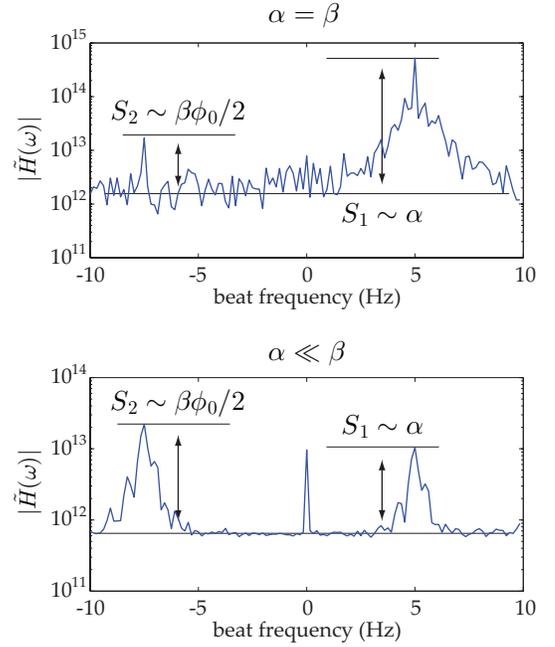}
\caption{Magnitude of the holograms $\tilde H (\omega)$ (a.u., vertical axis) versus beating frequency $\omega /(2\pi)$ (Hz, horizontal axis), averaged over the image of the piezo-electric actuator, for a vibration amplitude $z_0 \sim 30 \, \rm nm$. The top spectrum was acquired with a LO for which $\beta = \alpha$. For the bottom spectrum $\beta/\alpha = 50$.}
\label{fig_BeatFrequencySpectra}
\end{figure}

The proposed imaging apparatus is based on an optical interferometer of the same nature as for single mode heterodyne detection \cite{RoyerDieulesaint1986}, at the difference that a sensor array is used instead of a single photodiode (Fig.~\ref{fig_SetupSurfaceAcousticWaves}). The main laser beam (Cobolt Samba-TFB-150, linewidth $<1 \, \rm MHz$, wavelength $\lambda = 532 \, {\rm nm}$, optical frequency $\omega_{\rm L}/(2 \pi) = 5.6 \times 10^{14} \, \rm Hz$) is split into two channels. In the probe channel, the optical field $E$ is backscattered by the object, a thin metal plate, in vibration at the angular frequency $\omega$. For vibration amplitudes smaller than the optical wavelength, two modulation sidebands at $\omega_{\rm{L}} \pm \omega$ arise. In the reference channel, the LO optical field $E_{\rm LO}$ is frequency-shifted by two acousto-optic modulators (AA Opto-Electronic MT200-BG9-532-FIO-SM-J1-S-s) from which alternate diffraction orders ($\pm1$) are selected. The sensor array of the camera (Andor IXON 885+, 1002 $\times$ 1004 pixels of size $d = 8 \, \mu \rm m$, quantum efficiency $\eta \sim 0.55$) records the interference pattern of both optical fields beating against each other, in time-averaging conditions, at a sampling frame rate $\omega_{\rm S}/(2 \pi) = 20 \, \rm Hz$.  The distance between the plate and the sensor array is $\sim$ 50 cm. The phase modulation of the object field $E$ at angular frequency $\omega$ results in the apparition of optical sidebands of complex amplitude ${\cal E} J_n \left(\phi_0\right)$ at harmonics of $\omega$; for small modulation depths, the magnitude of the sidebands of order $\pm 1$ is much greater than the magnitude of the sidebands of higher order, as reported in figure \ref{fig_BeatFrequencySpectra}(a). The modulation depth of the optical phase is $\phi_0 = 4\pi z_0/\lambda$. The temporal part of the backscattered object field $E$ undergoing a sinusoidal optical path length and hence phase modulation $\phi(t) = 4\pi z(t) / \lambda = \phi_{0} \sin(\omega t)$ at the angular frequency $\omega$, can be derived from the first-order Taylor development of the modulated phase factor with respect to $\phi_0$
\begin{equation}\label{eq_TaylorPhaseModulationSine}
E(t) \approx {\cal E} e^{i\omega_{\rm{L}}t} \left( 1 + \frac{1}{2} \phi_0 e^{i\omega t} - \frac{1}{2} \phi_0 e^{-i\omega t}  \right)
\end{equation}

\begin{figure}[]
\centering
\includegraphics[width = 8.0 cm]{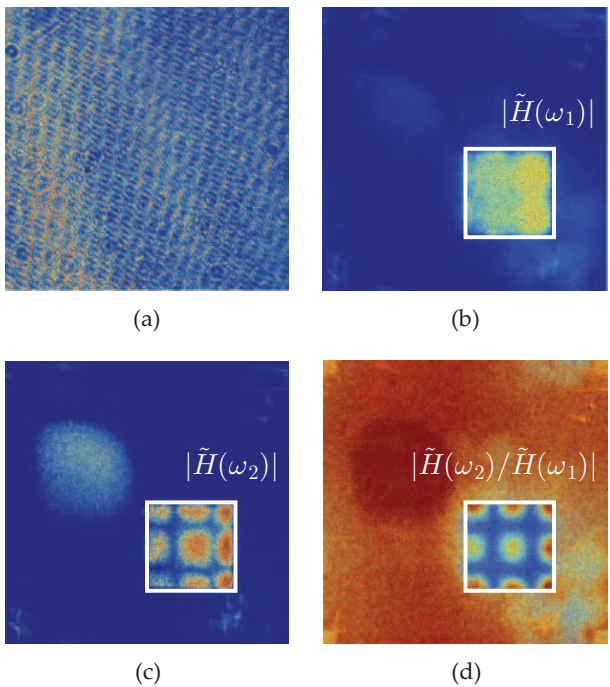}
\caption{Raw interferogram (a), hologram of statically-scattered light (central band), beating and $\omega_1$  (b), hologram of the first modulation sideband beating at $\omega_2$ (c), side-to-central band ratio (d). Colorbar in arbitrary logarithmic units.}
\label{fig_FourHolograms.eps}
\end{figure}

The LO signal consists of the addition of two coherent (phase-locked) radiofrequency (RF) signals, shifted by a carrier frequency $\omega_{\rm C}/(2\pi) \sim 200\, \rm  MHz$ set around the peak frequency response of acousto-optic modulators used to shift the optical frequency of the laser beam. This summation is done in practice with a power splitter/combiner (Fig.~\ref{fig_SetupSurfaceAcousticWaves}), resulting in a frequency-shifted optical LO field of the form
\begin{equation}\label{eq_DualLO}
E_{\rm LO}(t) = {\cal E}_{\rm LO} e^{i\omega_{\rm{L}}t} \left( \alpha e^{-i\omega_1 t} + \beta e^{i\omega t} e^{-i\omega_2 t} \right)
\end{equation}
where $\omega_1$ and $\omega_2$ are the intermediate frequencies chosen to shift the sidebands of order 0 and 1 in the beating-frequency spectrum of the recorded interferogram, at $\omega_1$ and $\omega_2$ respectively. The complex magnitudes of the LO components are ${\cal E}_{\rm{LO}_1} = \alpha {\cal E}_{\rm{LO}}$ and ${\cal E}_{\rm{LO}_2} = \beta {\cal E}_{\rm{LO}}$. The positive parameters $\alpha$ and $\beta$ are the relative weights of each LO field component, which satisfy $\alpha^2 + \beta^2 = 1$ (hence the total LO optical power scales up linearly with $\left|{\cal E}_{\rm LO}\right|^2$ and does not depend on $\alpha$ nor $\beta$). In high heterodyne gain regime, the optical power in the reference channel is much larger than in the object channel $\left|{\cal E}_{\rm{LO}}\right|^2 \gg \left|{\cal E}\right|^2$. The interference pattern (reported in Fig.~\ref{fig_FourHolograms.eps}(a)) holding RF fluctuations, from which optical frequencies are averaged-out, is
\begin{equation}\label{eq_InterferencePattern_E}
I(t) = |E + E_{\rm LO}|^2  
\end{equation}
The spatially-modulated component of the interferogram due to the off-axis configuration \cite{LeithUpatnieks1962} is $M E E^*_{\rm LO}$, where $M$ is the fringe visibility (the contrast of the interference fringes). Under the assumption that the excitation frequency is much larger than the frame rate, $\omega \gg \omega_{\rm S}$, the remaining terms within the temporal bandwidth of the sensor are
\begin{equation}\label{eq_TimeAveragedOffAxisComponent}
H(t) = \alpha M {\cal E}{\cal E}_{\rm LO}^* e^{-i\omega_1 t}  + \beta M \frac{\phi_0}{2}{\cal E}{\cal E}_{\rm LO}^* e^{-i\omega_2 t} + {\rm DC}
\end{equation}
where DC is the zero frequency component, observed in Fig.~\ref{fig_BeatFrequencySpectra} at 0 Hz. The off-axis hologram $H(t)$ is demodulated with a discrete Fourier transform (DFT) of $R$ consecutive interferograms \cite{VerrierAtlan2013, BrunoLaudereau2013}. The obtained beating frequency spectrum reveals the magnitude of the modulated components of Eq.~\ref{eq_TimeAveragedOffAxisComponent} between the Nyquist frequencies $\pm \omega_{\rm S} / 2$, with a spectral resolution $\omega_{\rm S}/R$. Hence time-averaging during the sequence of $R$ consecutive frames defines the width of the narrow band detection. The DFT $\tilde{H}(\omega)$ of $H(t)$, calculated over $R = 8$ frames, exhibits two peaks at the frequencies $\omega_1$ and $\omega_2$, reported in Fig.~\ref{fig_BeatFrequencySpectra}. The intermediate frequencies were set to non-opposite values $\omega_1 = \omega_{\rm{S}}/4$ and $\omega_2 = - 3\omega_{\rm{S}}/8$ in order to avoid cross-talk effects. The DFT's complex values yield holograms $\tilde{H}(\omega_1)$ (Fig.~\ref{fig_FourHolograms.eps}(b)) and $\tilde{H}(\omega_2 )$ (Fig.~\ref{fig_FourHolograms.eps}(c)), whose ratio (Fig.~\ref{fig_FourHolograms.eps}(d)) can be used to derive a quantitative map of the local out-of-plane vibration amplitude $z_0$
\begin{equation}\label{eq_RatioH}
\left| \frac{\tilde{H}(\omega_2 )}{ \tilde{H}(\omega_1 )} \right| = \frac{\beta}{\alpha} \frac{2 \pi}{\lambda} z_0
\end{equation}

\begin{figure}[b]
\centering
\includegraphics[width = 8.0 cm]{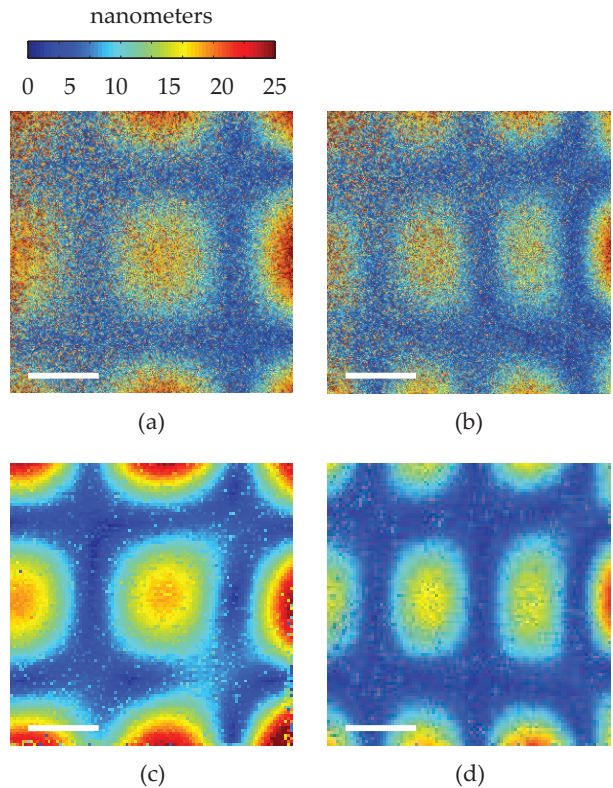}
\caption{
Amplitude maps of the out-of-plane vibration of a thin metal plate versus excitation frequency $\omega/(2\pi)$. Holographic images at 40.1 kHz (a), 61.7 kHz (b). Scanning laser Doppler images at 40.1 kHz (c), 61.7 kHz (d). Scalebar: 5 mm.}
\label{fig_SurfaceModes}
\end{figure}

A comparison of the contour modes of steady-state waves in a thin metal plate in sinusoidal vibration over $\sim$ 3 cm $\times$ 3 cm was performed by heterodyne holography against a single mode scanning laser Doppler probe \cite{RoyerDieulesaint1986, RoyerDieulesaint1989}. Vibration maps at two ultrasonic frequencies are reported in Fig.~\ref{fig_SurfaceModes}. For the single mode detection scheme, the scanning probe had an output power of 100 mW at a wavelength of 532 nm. It was focused and then swept over 100 $\times$ 100 points. The acquisition time per pixel was set to $1 \, {\rm ms}$ and averaged 512 times, leading to a signal acquisition time of $\sim 90 \, \rm min$ out of a total scanning time of about 3 hours per image. The contour modes obtained are reported in Fig.~\ref{fig_SurfaceModes}(c,d). The detector bandwidth was 40 MHz. For holography, the optical power over the whole object was about 30 mW and the distance from object to sensor was around 50 cm. Steady-state surface acoustic wave maps presented in Fig.~\ref{fig_SurfaceModes}(a,b) were recorded in 0.5 second with the holographic setup and the refreshment rate of the measurement was the video rate (20 Hz).

\begin{figure}[]
\centering
\includegraphics[width = 8.0 cm]{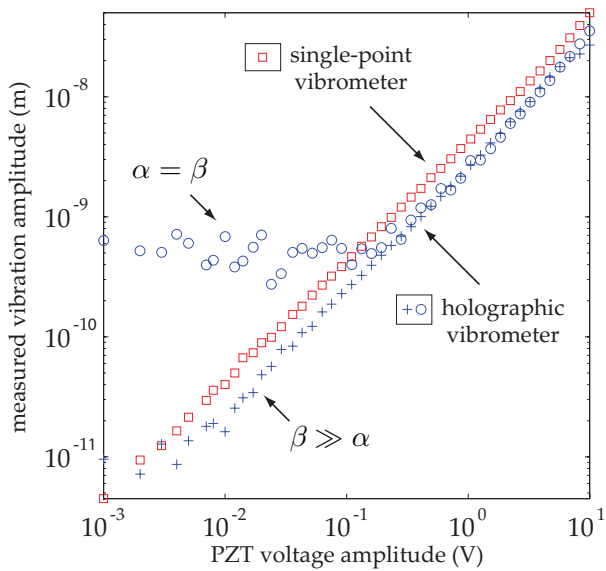}
\caption{Comparison of quantitative out-of-plane vibration amplitudes retrieved with the single-point laser vibrometer ($\square$ symbols) and holographic vibrometer with $\beta = \alpha$ ($\circ$ symbols) and $\beta \simeq 50 \, \alpha$ (+ symbols).}
\label{fig_CalibrationAmplitudeVersusVoltage}
\end{figure}

The linearity, sensitivity and dynamic range of the heterodyne holographic sensor was calibrated against the single mode laser Doppler probe \cite{RoyerDieulesaint1986, RoyerDieulesaint1989, BarriereRoyer2001}. We sought to compare vibration amplitudes derived from both methods, measured on the surface of the same piezo-electric actuator whose sinusoidal voltage at a frequency of 50 kHz was swept from 1 mV to 10 V to assess vibration amplitudes from $\sim$ 10 pm to $\sim$ 100 nm. Good agreement was found, as reported in Fig.~\ref{fig_CalibrationAmplitudeVersusVoltage}. Holographic vibrometry was assessed for two different values of the parameters  $\alpha$ and $\beta$: (i) $\beta = \alpha$ and (ii) $\beta = 50 \, \alpha$. The scanning laser Doppler probe and the holographic probe kept the same characteristics as formerly.

From a radiometric point of view, the average number of photons $n$ recorded per pixel of size $d$ during the measurement time $T$ satisfies $n \hbar \omega_{\rm L} =  |{\cal E}|^2 \epsilon_0 c d^2 T/ 2$, where $\hbar$ is the reduced Planck constant and $c$ is the speed of light. The average number of photons of the LO satisfies $n_{\rm LO} \hbar \omega_{\rm L} = |{\cal E}_{\rm LO}|^2 \epsilon_0 c d^2 T/ 2$.  From the relationship between the modulation amplitude and the sideband magnitude reported in Eq.~\eqref{eq_TaylorPhaseModulationSine}, the average magnitude of $\tilde{H}(\omega_1)$ and $\tilde{H}(\omega_2)$, in number of photo-electrons per pixel recorded during the measurement time $T$, for small vibration amplitudes $z_0\ll \lambda$, is $S_1 \approx \alpha M \eta (n n_{\rm LO})^{1/2}$ and $S_2 \approx \beta M  \eta (n n_{\rm LO})^{1/2} \phi_0 /2$, respectively, as illustrated in Fig.~\ref{fig_BeatFrequencySpectra}. The cumulative squared magnitude $S_2^2$ of the first sideband over the $N_{\rm px}$ pixels, in object beam-generated photo-electrons $n_e = \eta n$ recorded during the measurement time $T$ is
\begin{equation}\label{eq_SquaredMagnitudeHeterodyneSignal_n}
S_2^2 \approx \beta^2 N_{\rm px}^2 M^2 \eta ^2 n n_{\rm LO} \phi_0^2/4
\end{equation}
The cumulative energy in the zero-order sideband is 
\begin{equation}\label{eq_SquaredMagnitudeHeterodyneSignal_n_zero}
S_1^2 \approx \alpha^2 N_{\rm px}^2 M^2 \eta ^2 n n_{\rm LO}
\end{equation}
Under practical conditions, the detection is shot-noise limited in low light \cite{GrossAtlan2007}. The shot noise, in high heterodyne gain regime ($n_{\rm LO} \gg n$), induces an average photo-electrons current during time $T$ of  $\sqrt{ 2 e B \times 2 e B \eta n_{\rm LO} }$ per pixel \cite{WhitmanKorpel1969, UedaMiida1976, Monchalin1985, WagnerSpicer1987, RoyerDieulesaint1986}. The cumulative shot-noise variance $N^2$ over the $N_{\rm px}$ pixels, in photo-electrons, during the measurement time $T = 2 \pi R / \omega_{\rm S}$, equivalent to the Nyquist bandwidth $B = (2T)^{-1}$, is
\begin{equation}\label{eq_ShotNoiseVariance_n}
N^2 \approx N_{\rm px} \eta  n_{\rm LO}
\end{equation}
The sensitivity limit can be defined as the minimal vibration amplitude for which the signal-to-noise ratio of the magnitude of the first sideband is equal to one. Using relationships \eqref{eq_SquaredMagnitudeHeterodyneSignal_n} and \eqref{eq_ShotNoiseVariance_n}, this limit is
\begin{equation}\label{eq_SNR_n}
S_2^2 / N^2 = 1 \approx N_{\rm px} M^2 \beta^2 \eta n \phi_0^2/4
\end{equation}
The repartition of the sidebands energy in the detection can be tuned by $\alpha$ and $\beta$. For efficient experimental sensing of small-amplitude vibrations, the parameter $\beta$ has to be as high as possible but at the same time, $S_1^2$ has to remain at least equal to $S_2^2$, leading to the equality of the cumulative energy in the two sidebands and $\beta \Phi_0/2 = \alpha$. in addition to the energy conservation relationship $\alpha^2 + \beta^2 = 1$, we obtain that $\beta = 1$ and $\alpha \approx \Phi_0/2$. Hence the minimum out-of-plane vibration amplitude $z_{\rm min}$ that can be measured from a signal beam energy $n \hbar \omega_{\rm L} $ impinging on the sensor during time $T = 2 \pi N_{\rm img} / \omega_{\rm S}$, averaged over $N_{\rm px}$ pixels is
\begin{equation}\label{eq_ResolutionBruno}
z_{\rm min} = \frac{\lambda}{2 \pi} \frac{1}{M}  \frac{1}{\sqrt{ N_{\rm px} n_e}} = 2.9 \times 10^{-13} \, {\rm m}
\end{equation}
We measured a fringe visibility of $M\simeq 0.3$ on raw interferograms and estimated the number of photons per pixel recorded during the exposure time to be $n_e/\eta \sim 1.7 \times 10^6$, with $\eta n = n_e$, where $n_e$ is the number of photo-electrons created by the optical signal wave impinging on one sensor during the measurement time. In comparison, the absolute sensitivity of single mode optical heterodyne detection schemes \cite{WhitmanKorpel1969, UedaMiida1976, Monchalin1985, WagnerSpicer1987, RoyerDieulesaint1989, Kokkonen2008}, can theoretically reach $\sim 2.3 \times 10^{-15}$ m for a detection bandwidth of 1 Hz and a detected probe beam of 1 mW. Nevertheless, holographic vibrometry provides much more versatility than scanning interferometry in practical conditions for whole field monitoring of surface acoustic waves. It enables long-range imaging of rough surfaces at low optical power in real-time.

In conclusion, we demonstrated video-rate imaging of steady-state surface acoustic waves with a laser Doppler vibrometer based on optical holographic detection with a sensor array. A high temporal coherence laser was used to illuminate in wide field a thin metal plate in vibration. Narrow band recording of the map of out-of-plane vibration amplitudes was performed by heterodyne optical detection in time-averaging conditions with a 20 Hz frame rate camera. The high sensitivity of the measurement enabled amplitude retrieval from 100 nm down to 10 pm, which was calibrated against a standard laser Doppler vibrometer. The accuracy and sensitivity of the measurement is obtained by a detection process involving both spatial and temporal modulation of the interference pattern through off-axis and frequency-shifting holography, as well as a frequency-division multiplexing scheme, which was used to ensure simultaneous measurement of two modulation sidebands at distinct beating frequencies of the recorded interferogram. Robust and quantitative narrow band imaging of surface acoustic waves involved pixel-to-pixel division of two sideband holograms. In this regime, holographic vibrometry has the advantage of being self-calibrated: quantitative maps of the local vibration amplitude can be measured without prior calibration.

We gratefully acknowledge support from Fondation Pierre-Gilles de Gennes (FPGG014), Agence Nationale de la Recherche (ANR-09-JCJC-0113, ANR-11-EMMA-046), r\'egion \^Ile-de-France (C'Nano, AIMA), and the "investments for the future" program (LabEx WIFI: ANR-10-LABX-24, ANR-10-IDEX-0001-02 PSL*).


\end{document}